\documentclass[prb,
superscriptaddress,showpacs,amsmath,amssymb]{revtex4}

\usepackage{graphicx}

\begin{document}

\title{On chiral magnetic effect in Weyl superfluid $^3$He-A}

\author{G.E.~Volovik}
\affiliation{Low Temperature Laboratory, Aalto University,  P.O. Box 15100, FI-00076 Aalto, Finland}
\affiliation{Landau Institute for Theoretical Physics RAS, Kosygina 2, 119334 Moscow, Russia}

\date{\today}

\begin{abstract}{ 
In the theory of the chiral anomaly in relativistic quantum field theories (RQFT) some results depend on regularization scheme at ultraviolet. In the chiral superfluid $^3$He-A, which contains two Weyl points and also experiences the effects of chiral anomaly, the "trans-Planckian" physics is known and the results can be obtained without  regularization. We discuss this on example of the chiral magnetic effect (CME), which has been observed in $^3$He-A in 90's.\cite{Krusius1998}  There are two forms of the contribution of the CME to the Chern-Simons term in free energy, perturbative and non-perturbative. The perturbative term comes from the fermions living in the vicinity of the Weyl point, where the fermions are "relativistic" and obey the Weyl equation. The non-perturbative term originates from the deep vacuum, being determined by the separation of the two Weyl points in momentum space. Both terms are obtained using the Adler-Bell-Jackiw equation for chiral anomaly, and both agree with the results of the microscopic calculations in the  "trans-Planckian" region.
Existence of the two nonequivalent forms of the Chern-Simons term demonstrates that the results obtained within the RQFT depend on the specific properties of the underlying quantum vacuum and may reflect different physical phenomena in the same vacuum. 
}
\end{abstract}

\maketitle

\section{Introduction}

In relativistic quantum field theories (RQFT) many results crucially depend on the regularization scheme
in the high-energy corner (ultraviolet). This especially concerns the effect of chiral anomaly, see e.g. review \cite{Landsteiner2013} and discussion of consistent anomaly vs conventional anomaly in Ref.\cite{Stone2016}. The results may differ for example by the factor 1/3. 
We discuss here the effects of chiral anomaly which exist and have been experimentally observed in chiral Weyl superfluid $^3$He-A.

Superfluid $^3$He-A has two well separated Weyl points in the fermionic spectrum. Close to the Weyl points, quasiparticles obey the Weyl equation and behave as Weyl fermions, which interact with emergent gauge and gravity fields, see book \cite{Volovik2003} and references therein, and also recent review on topological superfluids \cite{Mizushima2016}. The dynamics of $^3$He-A is accompanied  by the hydrodynamic anomalies which reflect the chiral anomaly n the Weyl superfluids. The anomaly is described by the 
Adler-Bell-Jackiw equation\cite{Adler1969,Adler2005,BellJackiw1969},
which has been experimentally verified in experiments 
with skyrmions in $^3$He-A \cite{BevanNature1997}. The anomaly also gives rise to the analog of chiral magnetic effect (CME), which has been also experimentally identified in $^3$He-A 
\cite{Magnetogenesis1996,ExperimentalMagnetogenesis1997,Krusius1998,Volovik1998}.
The CME is the appearance  of the non-dissipative current along the magnetic field due to the chirality imbalance
\cite{Vilenkin1979,Vilenkin1980}. It is now under investigation in relativistic heavy ion collisions, where strong
magnetic fields are created by the colliding ions, see review \cite{Kharzeev2015}.
The theory of CME also experiences the problems, such as with the choice of the proper cut-off, see e.g.  Ref. \cite{Zubkov2016},
and the choice of the proper order of limits in the infrared \cite{Yamamoto2015}.

In $^3$He-A the hydrodynamic anomalies can be calculated either using the equations of the microscopic  theory (analog of the trans-Planckian physics), or using  the spectral flow described by the effective RQFT which emerges in the vicinity  of the Weyl points. The comparison between the two approaches provides the independent confirmation of the results obtained in the RQFT corner. The isolated Weyl points  are topologically  protected and survive when the interaction between quasiparticles is taken into account.  The integer valued topological invariant for the Weyl point is determined as the magnetic charge  of the Berry phase monopole in momentum space \cite{Volovik1987,Volovik2003}. 
The expansion of the Bogoliubov-de Gennes Hamiltonian  in the vicinity of the Weyl point produces the following  relativistic Hamiltonian describing the chiral Weyl fermions: 
\begin{equation}
H({\bf k})=e_\alpha^i\tau^\alpha \left(p_i-qA_i\right) \,.
\label{SU2}
\end{equation}
Here 
${\mbox{\boldmath$\tau$}}$ are Pauli matrices corresponding to the Bogoliubov-Nambu  isotopic spin; $q{\bf A}$ determine positions of two Weyl points ${\bf K}_\pm=q{\bf A}$, where the effective gauge field is expressed in terms of the direction of the orbital angular momentum of Cooper pairs, ${\bf A}({\bf r})= k_F\hat{\bf l}({\bf r})$, and $q=\pm 1$  is the effective electric charge  of the Weyl fermions living in the vicinity of two points.
The matrix $e_\alpha^i$ describes the effective (synthetic) tetrad field with ${\bf e}_1= c_\perp\hat{\bf m}$, ${\bf e}_2=c_\perp\hat{\bf n}$ and
${\bf e}_3=\pm c_\parallel\hat{\bf l}$, where $c_\perp$ and $c_\parallel$ are the components of effective speed of light;  $\hat{\bf m}$, $\hat{\bf n}$ and $\hat{\bf l}$ are orthogonal unit vectors. The determinant of the matrix determines the chirality of the Weyl fermions.  

 When $k_F\rightarrow 0$, the Weyl points merge and annihilate, as a results the hydrodynamic anomalies disappear. They do not exist in the strong coupling regime, where the chiral superfluid has no Weyl points.  All the dynamic anomalies experienced by $^3$He-A result from the existence of the Weyl points, which allow the spectral flow from the vacuum state of the superfluid. The state of the chiral superfluid with anomalies and the anomaly-free state of the chiral superfluid are separated by the topological Lifshitz transition (on topological quantum phase transitions 
see reviews\cite{Volovik2007,Volovik2017}).

\section{Chiral anomaly}

The chiral fermions experience the effect of chiral anomaly in the presence of the synthetic 
electric and magnetic fields 
\begin{equation}
{\bf A}=k_F\hat{\bf l} ~~,~~
{\bf E}=-\partial_t(k_F\hat{\bf l}) ~~,~~{\bf B}=\nabla \times (k_F\hat{\bf l} )
\,.
\label{EMfields}
\end{equation}
The fermions created from the superfluid vacuum carry the fermionic charge from the vacuum to the "matter" -- the normal component of the liquid, which at low temperatures consists of thermal Weyl fermions. For the chiral anomaly in $^3$He-A  the relevant fermionic charge is the quasiparticle momentum: each fermion created from the vacuum near the Weyl points at ${\bf K}_\pm = \pm k_F\hat{\bf l}$ carries with it the momentum $\pm k_F\hat{\bf l}$. Since the chiralities of two Weyl points are opposite, the momenta created at these points sum up. Thus the Adler-Bell-Jackiw equation for anomaly gives the following momentum creation from the vacuum state of $^3$He-A per unit time per unit volume:
\begin{equation}
\dot{\bf P} =\frac{1}{2\pi^2}  k_F \hat{\bf l}\,({\bf B}  \cdot {\bf E})   
\,.
\label{MomentumProductionGeneral}
\end{equation}

\section{CME term in free energy}

In this section we consider how the axial anomaly equation (\ref{MomentumProductionGeneral}) gives rise to the anomalous term in the hydrodynamic energy functional -- the Chern-Simons term. For that we consider the following process of the spectral flow.
We choose the spacetime dependence of effective gauge field in such a way that $k_F$ depends only on time,  $k_F(t)$, while the orbital unit vector depends only on space,  $\hat{\bf l}({\bf r})$. We consider the process in which $k_F(t)$ changes from zero, where the anomaly is absent, 
to the final value.
Then the effective electric field ${\bf E}=-\hat{\bf l}\,\dot k_F$, and  the final momentum density is
\begin{equation}
 {\bf P} =-\frac{1}{2\pi^2}  \int dt \,\dot k_F \,k_F^2 \,\hat{\bf l}\,( \hat{\bf l}  \cdot \nabla \times \hat{\bf l} )  =
-\frac{k_F^3 }{6\pi^2}  \,\hat{\bf l}\,( \hat{\bf l}  \cdot \nabla \times \hat{\bf l} ) 
\,.
\label{FinalMomentum}
\end{equation}
This is the same anomalous term in the current density, which has been obtained by Cross\cite{Cross1975} using the gradient expansion of the free energy and supercurrents starting with the microscopic Hamiltonian in a weak coupling approximation.
 The main contributions come from the region of momenta far away from the nodes, i.e. in our notations at the trans-Planckian scales, where the relativistic physics and its anomalies are not applicable. Nevertheless, the Cross result has been reproduced using the relativistic physics at the Weyl points. The reason for that is that the spectral flow does not depend on energy and is the same near the Weyl points and far away from them.
This also provides the independent check of the validity of Eq.(\ref{MomentumProductionGeneral}). Earlier the Eq.(\ref{MomentumProductionGeneral}) has been obtained in Refs. \cite{VolovikMineev1981,VolovikMineev1982} using the results of the microscopic theory.

In the presence of superflow with velocity  ${\bf v}_{\rm s}$, the current in Eq.(\ref{FinalMomentum}) gives the following contribution to the free energy density, which is also consistent  with the results by Cross\cite{Cross1975}: 
\begin{equation}
f_{CS}^{(1)}= {\bf v}_{\rm s} \cdot {\bf P} =
-\frac{k_F^3 }{6\pi^2}  \,  ({\bf v}_{\rm s} \cdot \hat{\bf l})\,( \hat{\bf l}  \cdot \nabla \times \hat{\bf l} ) 
\,.
\label{fCS1}
\end{equation}

Now let us take into account that in the analogy with RQFT the superflow velocity plays the role of the chiral chemical potential.\cite{Volovik2003}  Experimentally the effective imbalance between the chiral chemical potentials of the left-handed and right-handed Weyl fermions is provided by the counterflow due to the Doppler shift: 
$\mu_{R,L}=\pm k_F \hat{\bf l}\cdot ({\bf v}_{\rm s}-{\bf v}_{\rm n})$. Here ${\bf v}_{\rm n}$ is the velocity of the normal component of the liquid. We shall consider the heat bath reference frame, where ${\bf v}_{\rm n}=0$. 
In this frame
\begin{equation}
2k_F({\bf v}_{\rm s} \cdot \hat{\bf l}) =\mu_R -\mu_L
\,.
\label{ChiralChemical}
\end{equation}
Then Eq.(\ref{fCS1}) can be fully expressed  in terms of the effective relativistic fields, and it becomes the Chern-Simons term in the energy functional:
\begin{equation}
 F_{CS}^{(1)}=  
\frac{1}{12\pi^2}  \,  (\mu_R -\mu_L) \int d^3r\,\,({\bf A}  \cdot {\bf B})  
\,.
\label{FCS1}
\end{equation}
In the RQFT the variation of energy (\ref{FCS1}) over the vector potential ${\bf A}$ results in the equilibrium current along magnetic field ${\bf B}$, i.e. in the chiral magnetic effect. 

\section{CME in helical instability}

The experimental signature of the CME in $^3$He-A is the helical instability of the superflow generated by the  Chern-Simons  term, which contains the helicity ${\bf A}  \cdot {\bf B}$ of the effective gauge field.
To study the helical instability, we need the quadratic form of energy in terms of perturbations $\delta \hat{\bf l}$ of the homogeneous flow state. For that one should consider another route of deformation leading to spectral flow. Let us take $k_F={\rm const}$; the constant superfluid velocity ${\bf v}_{\rm s}\parallel \hat{\bf z}$; 
$\hat{\bf l}_0 = \hat{\bf z}$; and 
$\hat{\bf l}({\bf r},t) = \hat{\bf z}+ \delta \hat{\bf l}(z,t)$. Since $\hat{\bf l}(z,t)$ is unit vector, $\delta \hat{\bf l}(z,t)\perp \hat{\bf z}$, and we take $\delta \hat{\bf l}(z,t)={\bf m}(z)t$, where ${\bf m}(z)\perp \hat{\bf z}$. 
Then the final momentum is
\begin{equation}
 {\bf P} =-\frac{1}{2\pi^2}  \int dt \,t  \,k_F^3 \,\hat{\bf z}\,( {\bf m}  \cdot \nabla \times {\bf m} )  =
-\frac{k_F^3 }{4\pi^2}  \, \hat{\bf l}_0\,( \delta\hat{\bf l} \, \cdot \nabla \times \delta\hat{\bf l} ) 
\,.
\label{FinalMomentum2}
\end{equation}
The corresponding energy density in the presence of the superflow is
\begin{equation}
f_{CS}^{(2)}= {\bf v}_{\rm s} \cdot {\bf P} =
-\frac{k_F^3 }{4\pi^2}  \,  ({\bf v}_{\rm s} \cdot \hat{\bf l}_0)\,( \delta\hat{\bf l}  \cdot \nabla \times \delta\hat{\bf l} ) 
\,.
\label{fCS2}
\end{equation}
In terms of the effective fields this gives the following Chern-Simons term:
\begin{equation}
 F_{CS}^{(2)}=  
\frac{1}{8\pi^2}  \,  (\mu_R -\mu_L)\int d^3r\,( {\bf A} \cdot{\bf B})  
\,.
\label{FCS2}
\end{equation}
This term has been also independently derived from the equations describing the "trans-Planckian" dynamics  of the quantum liquid, see Ref. \cite{Volovik1998} (there is misprint in Eq.(42) of Ref.\cite{Volovik1998}, the prefactor should be 1/4 instead of 1/2). Note that Eq.(\ref{FCS2}) does not coincide with the variation of Eq.(\ref{FCS1}). In the RQFT approach the reason is the different behavior of the spectral flow for perturbative and non-perturbative deformations. In the microscopic "trans-Planckian" theory the nontrivial dependence of superfluid velocity ${\bf v}_{\rm s}$ on the texture of $\hat{\bf l}$-vector (the so-called Mermin-Ho relation \cite{MerminHo1976}) and the dynamics of the orbital angular momentum  combine to give the 3/2  factor \cite{Volovik1998}.

\section{Discussion}

%%%%%%%%%%%%%%%%%%%%%%%%%%%%%%%%%%%%%%%%%%%%%%%%%%%%%%%
%%%%%%%%%%%%%%%%%%%%%%%%%%%%%%%%%%%%%%%%%%%%%%%%%%%%%%%
\begin{figure}%[t]
\centerline{\includegraphics[width=1.0\linewidth]{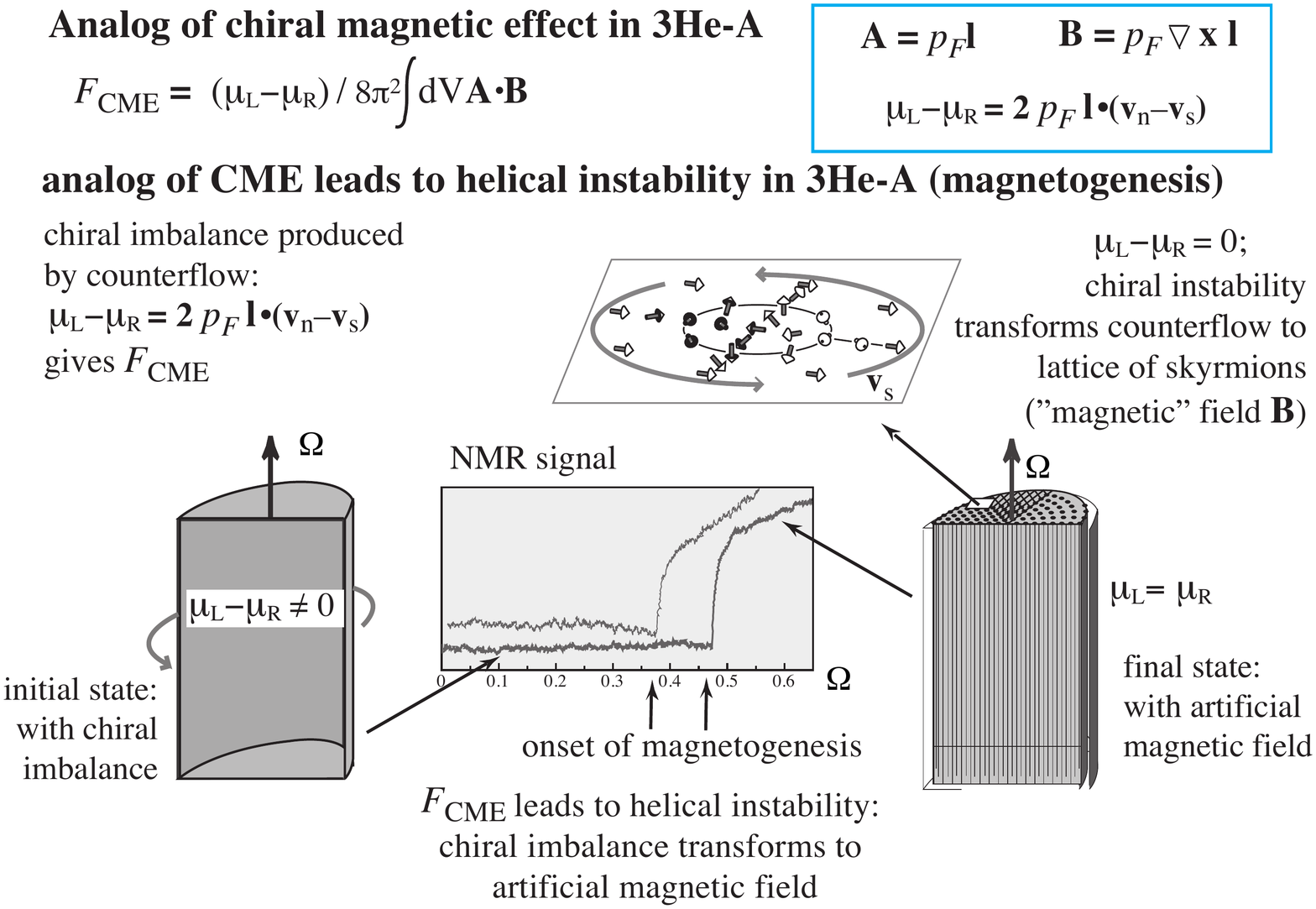}}
\medskip
\caption{CME in Weyl supefluid $^3$He-A. Counterflow (analog of chiral imbalance) experiences instability towards creation of vortex-skyrmions (analog of hypermagnetic field) due to the helical Chern-Simons energy term describing CME.\cite{Krusius1998} The experiment simulates the scenario of formation of primordial hypermagnetic field in early Universe due to axial (triangle) anomaly.\cite{JoyceShaposhnikov1997,GiovanniniShaposhnikov1998}
}
\label{CMEfigure}
\end{figure}
%%%%%%%%%%%%%%%%%%%%%%%%%%%%%%%%%%%%%%%%%%%%%%%%%%%%%%%
%%%%%%%%%%%%%%%%%%%%%%%%%%%%%%%%%%%%%%%%%%%%%%%%%%%%%%%

We obtained two expressions, Eq. (\ref{FCS1}) and Eq.  (\ref{FCS2}), for the Chern-Simons term emerging in superfluid $^3$He-A. For derivation we used the RQFT emerging in the vicinity of the Weyl points.  These two expressions differ by the factor 2/3. However, both terms are correct, since they coincide with the results of the microscopic nonrelativistic theory,  which is complete and thus does not require any regularization schemes. These two terms describe two different physical situations in the same underlying non-relativistic system. Eq. (\ref{FCS1})  represents the anomalous term in the hydrodynamic free energy, while the Eq. (\ref{FCS2}) determines the threshold of the helical instability of the superfluid flow.  This demonstrates that the results obtained within the RQFT may reflect the specific properties of the underlying quantum vacuum, and  thus can be different depending on what physical phenomena are considered in the same vacuum. 

In particular, even the existence of the  Chern-Simon term in energy depends of the behavior of the underlying vacuum. The variation of the Chern-Simon term over the real gauge potential, $\delta  F_{\rm CS}/\delta {\bf A}$, gives the current ${\bf J}$ along magnetic field ${\bf B}$. This is the signature of the
 chiral magnetic effect \cite{Vilenkin1979,Vilenkin1980}. However, if $U(1)$ symmetry is fully obeyed, the Bloch theorem prohibits the CME in the ground state of the system, see e.g. Ref.  \cite{Yamamoto2015}. This suggests that in this case the CME depends on the order
in which the limits $q\rightarrow 0$ and $\omega\rightarrow 0$ are taken \cite{Burkov2013}, and thus cannot be described by the Chern-Simon terms in bulk. 

In  case of $^3$He-A the gauge field ${\bf A} =  k_F\hat{\bf l}$ is effective. The emergent $U(1)$ symmetry exists only in the vicinity of the Weyl nodes, and not in the underlying quantum liquid. The corresponding current 
${\bf J}=\delta  F/\delta {\bf A}$ is also effective, and does not coincide with the real mass current 
${\bf P}=\delta  F/\delta {\bf v}_{\rm s}$. For the effective fields and currents the no-go Bloch theorem is not applicable, and the analog of the CME does exist. It has been experimentally 
confirmed in $^3$He-A \cite{Krusius1998,Magnetogenesis1996,ExperimentalMagnetogenesis1997} due to helical instability caused by the Chern-Simons helical term in Eq. (\ref{FCS2}). The chiral imbalance is provided by the superfluid flow, which is stable until the threshold is reached. Above the threshold the chiral imbalance becomes unstable towards production of the efective magnetic field (analog of magnetogenesis \cite{JoyceShaposhnikov1997,GiovanniniShaposhnikov1998}), see 
Fig. \ref{CMEfigure}.

I thank M. Zubkov for numerous discussions.
This project has received funding from the European Research Council (ERC)
under the European Union's Horizon 2020 research and innovation programme
(Grant Agreement \# 694248).

\end{document}